\documentclass[aps,prb,twocolumn,superscriptaddress,amsmath]{revtex4-1}

\usepackage[pdftex]{graphicx}

\begin{document}

\title{Effect of heavy-ion irradiation on London penetration depth in over-doped Ba(Fe$_{1-x}$Co$_x$)$_2$As$_2$}%

\author{J.~Murphy}
\affiliation{Ames Laboratory and Department of Physics and Astronomy, Iowa State University, Ames, Iowa 50011,
USA }

\author{M.~A.~Tanatar}
\affiliation{Ames Laboratory and Department of Physics and Astronomy, Iowa State University, Ames, Iowa 50011,
USA }

\author{Hyunsoo~Kim}
\affiliation{Ames Laboratory and Department of Physics and Astronomy, Iowa State University, Ames, Iowa 50011,
USA }

\author{W.~Kwok}
\affiliation{Argonne National Laboratory, Argonne, Illinois 60439, USA}

\author{U.~Welp}
\affiliation{Argonne National Laboratory, Argonne, Illinois 60439, USA}

\author{D.~Graf}
\affiliation{National High Magnetic Field Laboratory, Florida State university, Tallahassee, Florida 32310, USA}

\author{J.~S.~Brooks}
\affiliation{National High Magnetic Field Laboratory, Florida State university, Tallahassee, Florida 32310, USA}

\author{S.~L.~Bud'ko}
\affiliation{Ames Laboratory and Department of Physics and Astronomy, Iowa State University, Ames, Iowa 50011,
USA }

\author{P.~C.~Canfield}
\affiliation{Ames Laboratory and Department of Physics and Astronomy, Iowa State University, Ames, Iowa 50011,
USA }

\author{R.~Prozorov}
\affiliation{Ames Laboratory and Department of Physics and Astronomy, Iowa State University, Ames, Iowa 50011,
USA }

\date{\today}

\begin{abstract}
Irradiation with 1.4 GeV $^{208}$Pb ions was used to induce artificial disorder in single crystals of iron-arsenide superconductor Ba(Fe$_{1-x}$Co$_x$)$_2$As$_2$ and to study its effect on the temperature-dependent London penetration depth and transport properties. Study was undertaken on overdoped single crystals with $x$=0.108 and $x$=0.127 characterized by notable modulation of the superconducting gap. Irradiation with doses 2.22$\times$10$^{11}$$\textit{d}$/cm$^2$ and 2.4$\times$10$^{11}$$\textit{d}$/cm$^2$, corresponding to the matching fields of $B_{\phi} = $6~T and 6.5~T, respectively, suppresses the superconducting $T_c$ by approximately 0.3 to 1~K. The variation of the low-temperature penetration depth in both pristine and irradiated samples is well described by the power-law, $\Delta \lambda (T)=AT^n$. Irradiation increases the magnitude of the pre-factor $A$ and decreases the exponent $n$, similar to the effect of irradiation in optimally doped samples. This finding supports universal $s_{\pm}$ pairing in Ba(Fe$_{1-x}$Co$_x$)$_2$As$_2$ compounds for the whole Co doping range.

\end{abstract}

\maketitle

\section{Introduction}
Soon after discovery of superconductivity in iron-based materials \cite{Hosono}, it was recognized that the strength of electron-phonon coupling in the compounds is not sufficient to explain $T_c$ in 50~K range \cite{phonons}. Together with proximity to magnetic quantum critical point in the doping phase diagram \cite{Paglione,Johnston,CanfieldBudko,Stewart}, this fact is suggestive that superconductivity in iron-pnictides can be magnetically mediated, a scenario intensely discussed for cuprates, heavy fermion and organic superconductors \cite{Lonzarich,normanscience}.

Studies of the structure of the superconducting order parameter and thus pairing mechanism, can give an important insight into the problem. For the explanation of early experiments in iron pnictides, showing both full gap  \cite{ChenNature} and neutron resonance \cite{resonance}, a pairing state was suggested in which the superconducting order parameter changes sign on different Fermi surface sheets \cite{Mazinspm,MazinNature} and thus enables pairing by Coulomb repulsion \cite{Scalapino}. Contrary to the $d-$wave state in the cuprates \cite{dwave}, this so-called $s_{\pm}$ pairing may or may not exhibit nodes and if it does, the nodes are accidental \cite{HirschfeldReviewROPP}. The full gap is indeed found at optimal doping in electron-doped NdFeAs(O,F) \cite{ChenNature,Kondo} and Ba(Fe$_{1-x}$Co$_x$)$_2$As$_2$ (BaCo122 in the following) \cite{TanatarPRL,Reidcaxis}, hole-doped (Ba$_{1-x}$K$_x$)Fe$_2$As$_2$ (BaK122 in the following) \cite{ReidSUST}, and in stoichiometric LiFeAs \cite{HyunsooLiFeAs,ARPESLiFeAS,TanatarLiFeAs}. Studies of the doping evolution of the superconducting gap in BaCo122 and in BaK122 found a nearly universal development of a strong gap anisotropy and even nodes at the dome edges \cite{Fukazawa,Dong,HashimotoK,TanatarPRL,Reidcaxis,ReidK122PRL,ReidSUST}. An explanation of this evolution was suggested considering accidental nodes in the $s_{\pm}$ scenario as a result of a competition between intra-band Coulomb repulsion, tending to develop gap anisotropy within each band, and the inter-band attraction \cite{Chubukovreview} or as the phase transition from $s_{\pm}$ to a $d-$wave pairing state \cite{Thomale}. The former scenario is supported by the experimentally determined doping evolution of the gap in overdoped BaK122 \cite{BaKoverdopedARPES}, whereas the latter scenario finds support in the non-monotonic pressure dependence of $T_c$ \cite{Fasel} and universal character of thermal conductivity \cite{ReidK122PRL}.

While full isotropic superconducting gap at the optimal doping is, indeed, consistent with $s \pm$ model, this explanation is not unique. It was also discussed that orbital-mediated pairing can lead to a full gap $s_{++}$ state \cite{orbitalKOntani}. In addition, theoretical calculations show that $s_{\pm}$, $s_{++}$ and $d-$wave states are very close in energy, and the resultant ground state can depend sensibly on fine structural, magnetic and electronic details, - for example an angle of the As-Fe-As bond \cite{Kontani-angle}.

The experimental distinction between these possible pairing states is not trivial. It was suggested that because of the sign change of the order parameter in $s_{\pm}$, $T_c$ should be strongly sensitive to nonmagnetic impurities, similar to nodal $d-$wave, but the two states should show different evolution of the quasi-particle excitations. These can be revealed, for example, in the low-temperature exponents of the London penetration depth, $\Delta \lambda (T)=AT^n$, Ref. \onlinecite{Peterdoping}. Therefore, deliberate introduction of the additional disorder may serve as an important tuning parameter to distinguish between different superconducting states. Indeed, iron based superconductors have some inherent amount of disorder associated with random distribution of dopant atoms, required to induce superconductivity. This may explain experimentally observed power-law exponent $n$ in BaCo122, which is close to two for all dopings \cite{Gordondoping}.

It was known from studies on high-$T_c$ cuprates that extended defects, created by heavy-ion irradiation, not only act as efficient vortex pinning centers, but also as scattering sites. This is evident from a clear increase in normal state resistivity and a suppression of $T_c$. In iron arsenides irradiation with heavy ions indeed leads to the decrease of the exponent $n$ and increase of the pre-factor $A$ in both optimally doped BaCo122 \cite{hyunsooheavyion} and BaK122 \cite{Giannettaheavyion}, consistent with $s_{\pm}$ scenario. The change of $T_c$, however, is still an open question and is at best moderate \cite{hyunsooheavyion} or non-detectable \cite{Giannettaheavyion}. On the other hand, electron irradiation results in a clear reduction of $T_c$ in these systems, so disorder works as expected \cite{PrivateComm,vanderBeek}.

In this work we study experimentally the effect of heavy - ion irradiation in overdoped superconductors where the superconducting gap is strongly anisotropic from the start, even in the pristine state. Specifically, we measured London penetration depth (and thus quasi-particle excitations) in pristine and irradiated overdoped single crystals of BaCo122 family. These compositions are characterized by strong gap anisotropy as found in temperature and magnetic field response of thermal conductivity \cite{TanatarPRL,Reidcaxis}. The selection of these materials was motivated by the fact, that in $s_{++}$ superconductors with accidental nodes, the disorder should wipe out gap minima, and thus its effect should be very different from that of the $s_{\pm}$ state. Additionally, we studied resistivity in various magnetic fields and determined upper critical field, $H_{c2}$,  to test recently suggested link between the anisotropy of the gap and $T-$linear $H_{c2,c}(T)$ in $H \parallel c$ configuration \cite{SteveHc2,YLiu}.


\section{Experimental}

\subsection{Sample preparation}

Single crystals of BaFe$_2As_2$ doped with Co were grown from a starting load of elemental Ba, FeAs and CoAs, as described in detail elsewhere \cite{NiNiCo}. Crystals were thick platelets with sizes as big as 12$\times$8$\times$1 mm$^3$ and large faces corresponding to the tetragonal (001) plane.  The actual content of Co in the crystals was determined with wavelength dispersive electron probe microanalysis and is the $x$-value used throughout this text. The two compositions studied were $x$=0.108 ($T_c \approx$16~K) and $x$=0.127 ($T_c \approx$8~K), from the same batches used in previous thermal conductivity \cite{TanatarPRL,Reidcaxis} studies. They were on the overdoped side of the doping phase diagram (see inset in Fig.~\ref{resistivity}), notably above optimal doping level $x_{opt}$=0.07 ($T_c$ $\approx$23~K).

\subsection{Measurements of London penetration depth}

The in-plane London penetration depth was measured using the tunnel diode resonator (TDR) technique \cite{ProzorovSUST}.  The sample was mounted on a sapphire rod that was then inserted into the inductor coil (L) component of an LC Tank circuit, creating {\it ac} magnetic field, $H_{ac} \sim 20$ mOe. Since $H_{ac} \ll  H_{c1}$, the sample remains in the Meissner state and the magnetic response is governed by the London penetration depth. During experiments $H_{ac}$ was parallel to sample $c-$axis, thus measuring field penetration along the conducting plane.  The presence of the sample in the coil causes a frequency shift, which can be related to the change in the inductance of the TDR circuit $\Delta f=f_0-f(T)$ where $f_0=1/(2\pi \sqrt{LC})\sim 14$~MHz.  The real part of the magnetic susceptibility $\chi$(T) can then be derived $\Delta f=-G4 \pi \chi(T)$.  The calibration factor $G=f_0 V_s /(2 V_c (1-N)$ is defined by the sample volume, $V_s$, coil volume, $V_c$ and the demagnetization factor, $N$.  Experimentally $G$ is directly measured by physically pulling the sample out of the coil {\it in situ} at low temperatures.  With the sample in the Meissner state, $\lambda$ can be obtained from the following relation,  $4 \pi \chi (T)= \lambda/R \tanh{(R/\lambda)}-1$, where $R$ is the effective dimension of the sample \cite{Prozorov2000}.

To measure the changes in superconducting transition temperature, $T_c$, and the change in the penetration depth, $\Delta \lambda (T)$, with irradiation, the same samples were first  measured using a TDR setup in $^3$He cryostat (down to 0.5~K) and then in a dilution refrigerator down to 0.05~K \cite{KimSrPdGe}. Reference samples were stored in the same environment as irradiated samples, and re-measured to assure that there is no degradation during storage. They were used in the upper critical field measurements afterwards, with contacts soldered for resistivity measurements.

\subsection{Heavy ion irradiation}

To create columnar defects, samples were irradiated with 1.4GeV $^{208}$Pb ions at the Argonne Tandem Linear Accelerator System (ATLAS). The ions at this energy have stoppage distance of about 70 $\mu$m, so samples were cleaved to a thickness of 30 $\mu$m or less, to ensure homogeneous effect of irradiation. The flux and the dose of irradiation were measured during each irradiation experiment. The irradiation dose was 2.22$\times$10$^{11}$$\textit{d}$/cm$^2$ for $x$=0.127 sample and 2.4$\times$10$^{11}$$\textit{d}$/cm$^2$ for $x$=0.108 sample. Traditionally the density of columnar defects, $d$, is characterized using matching magnetic field, $B_{\phi} = \Phi_0 d$, calculated assuming one magnetic flux quantum $\Phi_0 \approx 2.07 \times 10^{-7}$ G$\cdot$cm$^2$ per ion track. The samples studied here where given 6 T and 6.5 T equivalent doses.

\subsection{Electrical resistivity and Upper Critical Field measurements}


\begin{figure}[tb]
\begin{center}
\includegraphics[width=8cm]{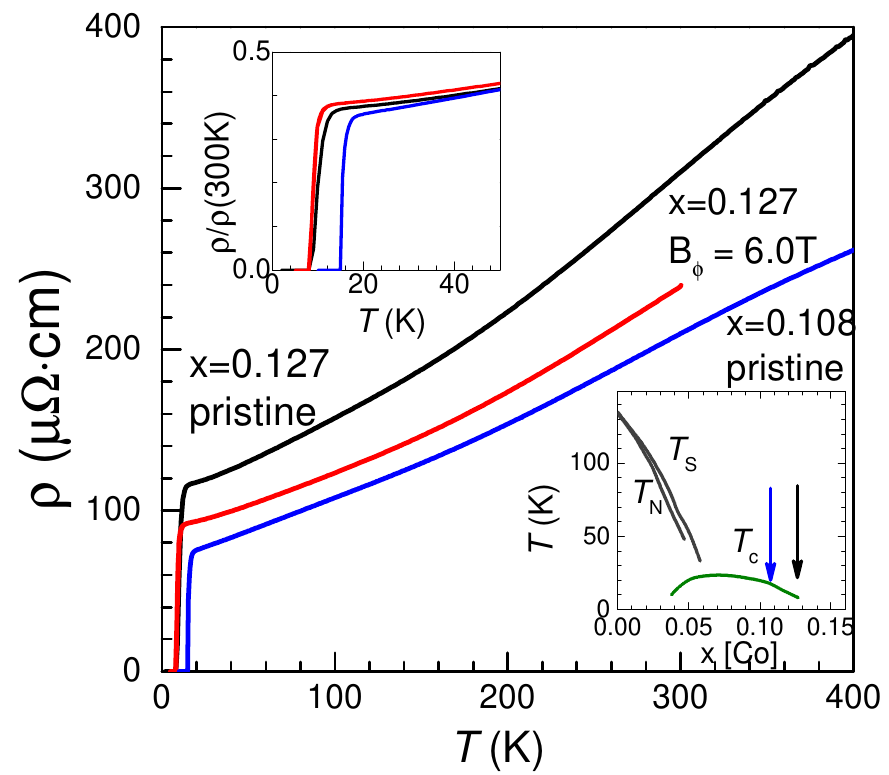}
\end{center}
\caption{(Color Online) Temperature dependent electrical resistivity of reference samples $x$=0.108 and $x$=0.127 and of the irradiated sample of $x$=0.127. The irradiated sample $x$=0.127 is the same sample as used in penetration depth measurements, with contacts soldered after measurements completed. Top inset shows same curves using normalized resistivity scale $\rho/\rho(300K)$, showing that the main difference between resistivity values comes from error of geometric factor determination \cite{NiNiCo,anisotropy}. Bottom inset shows the sketch of the doping phase diagram for BaCo122 with position of the samples used in this study. }%
\label{resistivity}
\end{figure}

In-plane electrical resistivity, $\rho$, and the upper critical field measurements were performed on reference samples cut from the same crystals as used in the penetration depth study. Samples were cleaved into rectangular shape, with crystallographic $a-$axis along the long side.  Contacts were made by soldering silver wires with ultrapure tin, \cite{SUST,PATENT} resulting in very low contact resistance (less than 10$\mu$$\Omega$).  Resistivity measurements were made using standard four-probe technique and samples were initially characterized using a {\it Quantum Design PPMS}. 
The temperature-dependent resistivity of these samples is shown in the main panel of Fig.~\ref{resistivity}.

To enable measurements in high magnetic fields and to prevent sample motion during in-field rotation, the samples were glued with GE varnish to a G-10 sample stage. Sample resistance was checked after mounting and found to agree with the measurements in a free-standing state.  The stage was fitted into a single axis rotator of a 35 T DC magnet at the National High Magnetic Field Laboratory in Tallahassee, Florida.  The rotator allows over 90$^0$ rotation around its horizontal axis in the vertical magnetic field.  Measurements were made in a $^4$He-cryostat with variable temperature insert (VTI) with lowest temperatures down to 1.5 K.  The rotator was equipped with a stepper motor with angular resolution of 0.01$^0$. The magnetic field was aligned parallel to the sample plane, $\theta$=0, using angular- dependent resistivity in a magnetic field slightly below H$_{c2,ab}$, see Ref.~\onlinecite{Murphy} for further details.

After finishing penetration depth study, four contacts were soldered \cite{SUST,PATENT} to the sample with $x$=0.127, and the temperature dependent resistivity $\rho(T)$ was measured, see Fig.~\ref{resistivity}. 
Contact soldering was made at $T \sim$~500~K, which could lead to a partial annealing of the irradiation damage. Note that the difference between pristine and irradiated samples of $x$=0.127 in Fig.~\ref{resistivity} mainly comes from the error in the geometric factor determination, particularly big in micaceous crystals of iron pnictides due to hidden cracks \cite{NiNiCo,anisotropy}. We failed to make contacts to a much thinner irradiated sample $x$=0.108.

\section{Results and Discussion}

\subsection{Resistivity change with irradiation}


\begin{figure}[tb]
\begin{center}
\includegraphics[width=8cm]{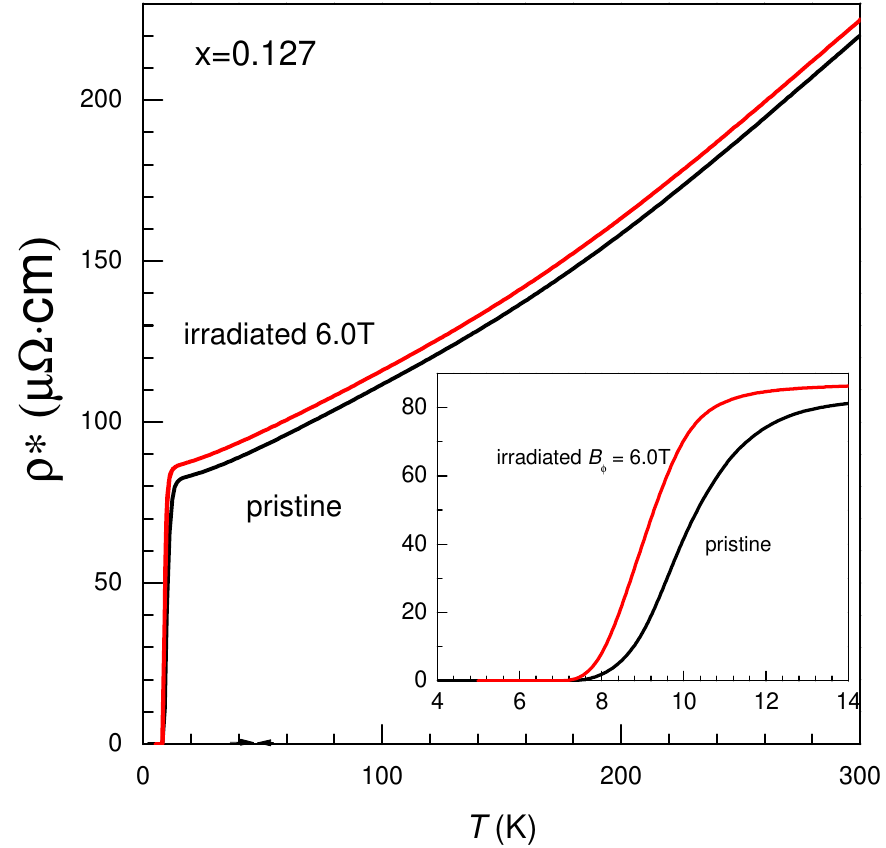}
\end{center}
\caption{(Color Online) Comparison of the temperature dependent electrical resistivity of reference and irradiated samples $x$=0.127, with geometric factor correction by normalizing slopes of the $\rho(T)$ curves at 300~K. For pristine sample we assumed $\rho(300K)$=220 $\mu \Omega$cm \cite{pseudogap}. Inset zooms at the superconducting transition range. Heavy ion irradiation does not change the shape of the $\rho(T)$ curves but increases residual resistivity, sharpens the superconducting transition and shifts $T_c$ down.  }%
\label{rho0}
\end{figure}

The $\rho(T)$ of irradiated sample, Fig.~\ref{resistivity}, is nearly the same within uncertainty of the geometric factor with that of the reference sample \cite{NiNiCo,anisotropy}. A rough way of removing this uncertainty is to normalize the resistivity by its value at the room temperature, $\rho (300K)$, as shown in top inset in Fig.~\ref{resistivity}. It reveals that the irradiated sample has higher normalized resistivity value at $T_c$ and slightly lower $T_c$, as indeed expected. To make a more careful $\rho(T)$ comparison, we normalized the slopes of $\rho(T)$ curves at 300~K. For pristine sample we also used the value of the resistivity at room temperature, $\rho(300K)$=220 $\mu \Omega$cm, as determined by statistically significant measurements on a big array of samples \cite{pseudogap}. In Fig.~\ref{rho0} we compare temperature dependence of thus adjusted resistivity $\rho^{\star} (T)$ for pristine and irradiated samples with $x$=0.127. 

Heavy ion irradiation has three effects on the $\rho^{\star}(T)$ of the samples. (1) Irradiation slightly parallel-up-shifts the $\rho(T)$ curve, as expected for samples with increased residual resistivity obeying Matthiessen's rule. The shift allows us to quantify additional increase of residual resistivity due to irradiation damage as $\Delta \rho^{\star}(0)$=4~$\mu \Omega$cm. This increase is significantly smaller that the extrapolated residual resistivity, $\rho (0) \approx$80$\mu \Omega$cm, which explains very small shift of the superconducting $T_c$. (2) Irradiation sharpens the resistive transition, which is presumably a reflection of the increased pinning on columnar defects and breaking weak links in the samples \cite{pinning}, and possible suppression of vortex fluctuations. (3) Irradiation shifts the onset and the midpoint $T_c$ of the superconducting transition by almost 1~K. This value is somewhat higher than found in the penetration depth measurements, see Fig.~\ref{irradiationlinear} below, however, because of the transition sharpening, it strongly depends on the criterion used.
Observation of the comparable $T_c$ shift suggests that short-time heat treatment at $\sim$500~K during contact soldering does not lead to significant annealing of the defects induced by the heavy ion irradiation.

\subsection{Doping evolution of the temperature dependent $H_{c2}(T)$}


\begin{figure}
\begin{center}
\includegraphics[width=8cm]{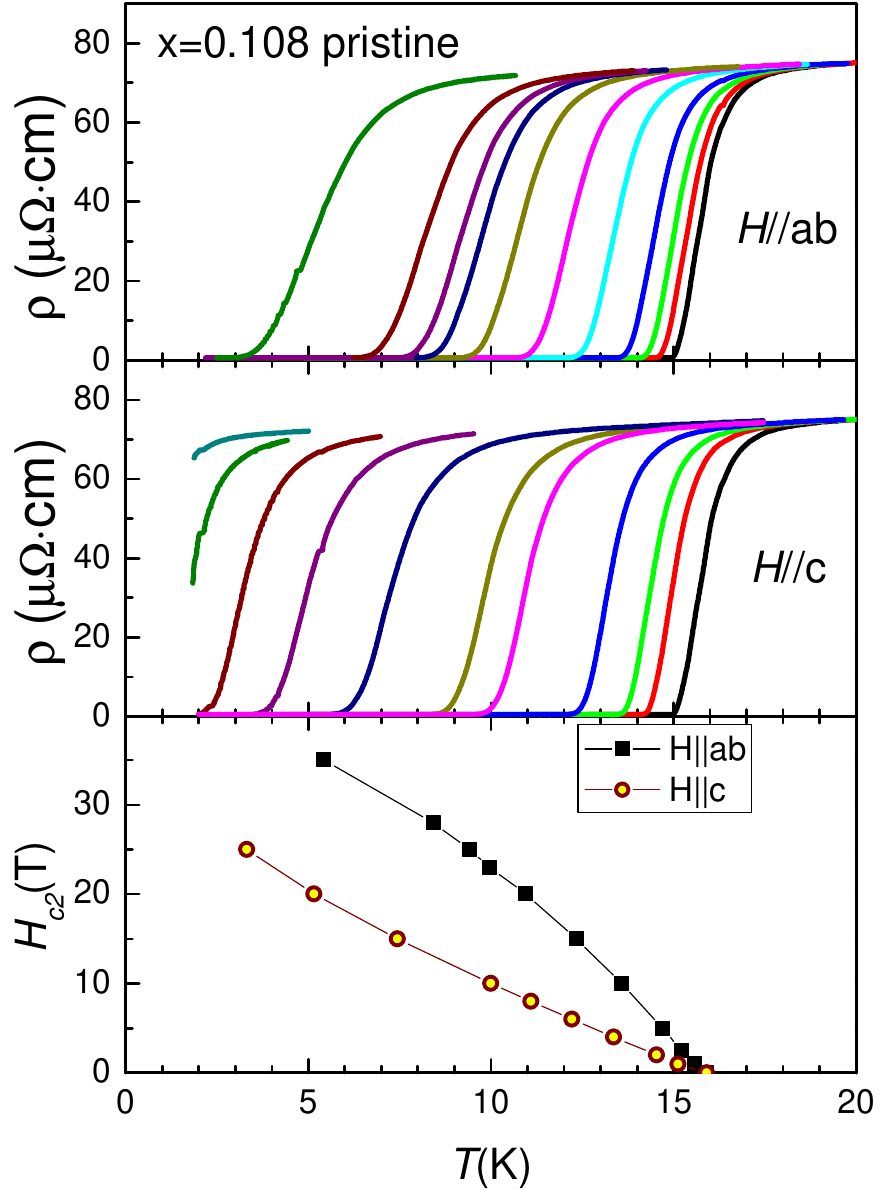}
\end{center}
\caption{(Color Online)  In-plane resistivity $\rho (T)$ for slightly overdoped Ba(Fe$_{1-x}$Co$_{x}$)$_2$As$_2$, $x$=0.108, in magnetic fields (a) parallel to the conducting plane, perpendicular to the $c$-axis and (b) parallel to the crystallographic $c$-axis. Field values (right to left) 0, 1, 2.5, 5, 10, 15, 20, 23, 25, 28, 35~T. Bottom panel (c) shows $H_{c2}$(T) phase diagrams for both directions of magnetic field. }%
\label{Hc2rawslight}
\end{figure}


\begin{figure}
\begin{center}
\includegraphics[width=8cm]{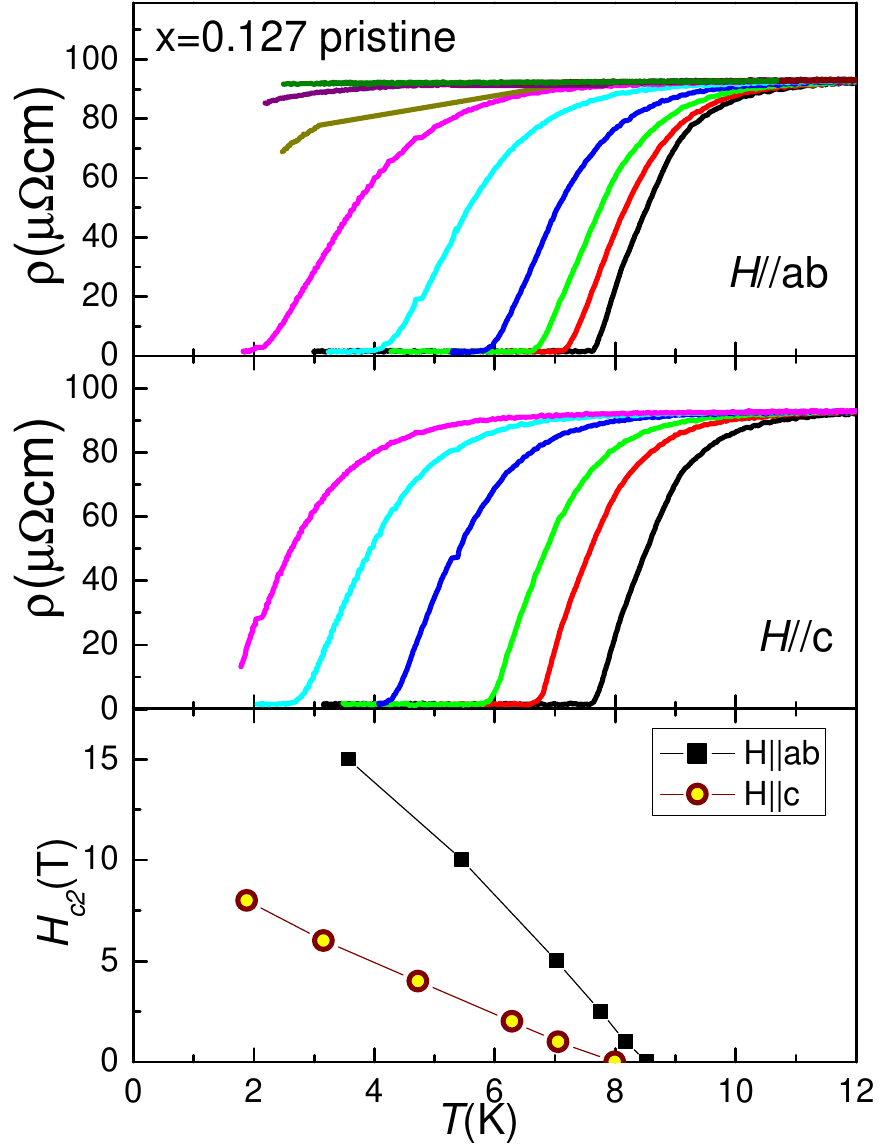}
\end{center}
\caption{(Color Online) In-plane resistivity $\rho (T)$ for heavily overdoped Ba(Fe$_{1-x}$Co$_{x}$)$_2$As$_2$, $x$=0.127 in magnetic fields (a) parallel to the conducting plane, perpendicular to the $c$-axis, field values (right to left) 0, 1, 2.5, 5, 10, 15, 20, 23 and 25~T.   and (b) parallel to the crystallographic $c$-axis, field values (right to left) 0, 1, 2, 4, 6, 8 and 10~T.   Bottom panel (c) shows $H_{c2}$(T) phase diagrams for both directions of magnetic field. }%
\label{Hc2rawstrong}
\end{figure}


\begin{figure}
\begin{center}
\includegraphics[width=8cm]{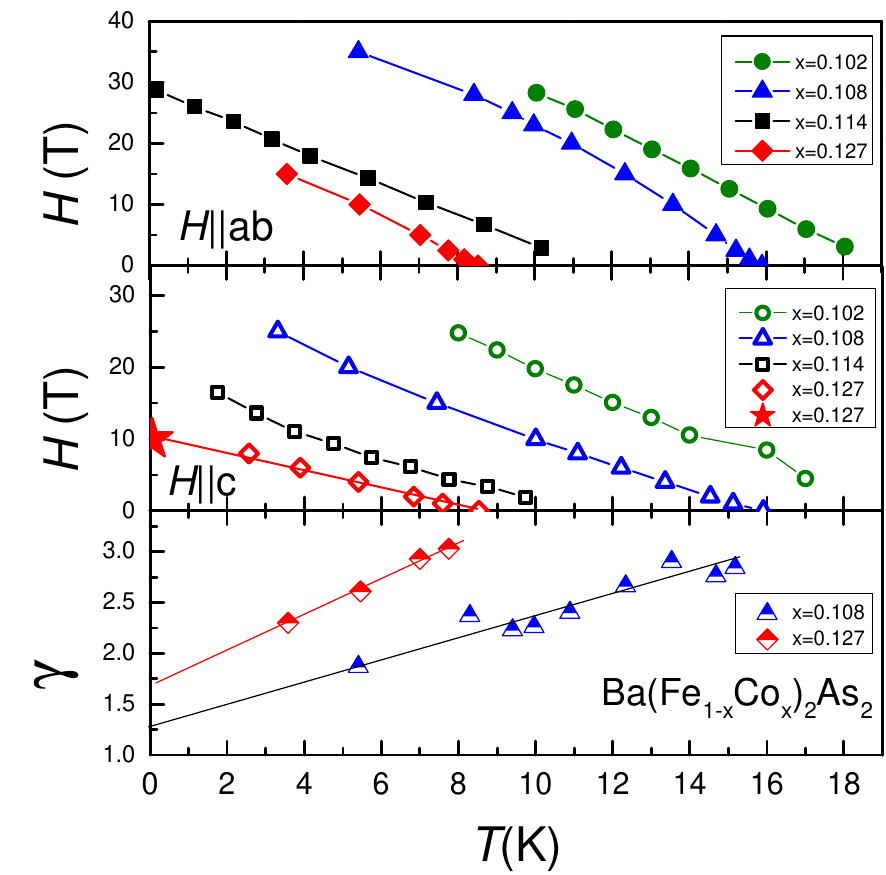}
\end{center}
\caption{(Color Online) $H_{c2}(T)$ and anisotropy parameter $\gamma \equiv H_{c2,ab}/H_{c2,c}$ for overdoped samples of BaCo122 $x$=0.108 and $x$=0.127. Top panel shows configuration with magnetic fields parallel to the conducting plane. Middle panel shows configuration with magnetic field parallel to the crystallographic $c$-axis. For reference we show bulk $H_{c2,c}(0)$ as determined from thermal conductivity study in sample $x$=0.127, Ref.~ \onlinecite{Reidcaxis}, which suggest that the data for strongly overdoped samples $x$=0.127 show very close to $T$-linear dependence. In top and middle panels we show data for previously studied overdoped samples with $x$=0.102 and $x$=0.114 \cite{NiNiCo}, determined using same mid-point criterion. Bottom panel shows temperature-dependent anisotropy parameter for samples $x$=0.108 and $x$=0.127. }
\label{Hc2Comparison}
\end{figure}


We reported recently that the shape of the temperature-dependent upper critical field for field orientation parallel to the tetragonal $c$-axis, $H_{c2,c}(T)$, is very different for iron-arsenide superconductors with full superconducting gap, (e.g., LiFeAs \cite{LiFeAsHc2}), and with nodal gap, (e.g., SrFe$_2$(As$_{1-x}$P$_x$)$_2$ \cite{SteveHc2,SrPpenetration} and KFe$_2$As$_2$ \cite{YLiu}). The former shows clear saturation at low temperatures, in line with expectations of both orbital WHH theory \cite{WHH}, and paramagnetic Clogston- Chandrasekhar limit \cite{CC}, while the latter remains close to $T-$linear down to the lowest temperatures.  Moreover, in KFe$_2$As$_2$ the low-temperature $H_{c2,c}(0)$ scales with $T_c$, which is not expected for standard orbital limiting mechanism in the clean limit \cite{YLiu,KoganProzorovROPP}.

To gain further insight into these unusual features we studied anisotropic upper critical fields in overdoped BaCo122 compositions, in which the superconducting gap anisotropy increases towards the superconducting dome edge. We used resistive $H_{c2}$ determination, in the constant-field temperature-sweep measurements, as shown in Figs.~\ref{Hc2rawslight} and \ref{Hc2rawstrong} for pristine samples of BaCo122 with $x$=0.108 and $x$=0.127, respectively, for field orientations parallel to the conducting $ab$ plane (top panels) and parallel to the tetragonal $c$-axis (bottom panels). A mid-point of the resistive transition was used as a criterion to determine $H_{c2}(T)$.
Bottom panels show $H-T$ phase diagrams determined from these measurements. 

In Fig.~\ref{Hc2Comparison} we compare the results of our measurements with the results of the previous study \cite{NiNiCo} on samples $x$=0.102 and $x$=0.114. The two sets are in good agreement and reveal very monotonic evolution of $H_{c2}$ with doping. Bottom panel in Fig.~\ref{Hc2Comparison} shows temperature-dependent anisotropy $\gamma \equiv H_{c2,ab}/H_{c2, c}$ for samples with $x$=0.108 and $x$=0.127. Close to $T_c$ the anisotropy is maximum, $\gamma$=2.5$\pm$0.5 for $x$=0.108 and $\gamma$=3.5$\pm$0.5 for $x$=0.127, with error bars determined by the difference in the criteria of resistive transition temperature determination. These values are consistent with the values found in the overdoped compositions in previous study and significantly different from much smaller anisotropies, $\gamma \sim$1, found in the underdoped compositions \cite{NiNiCo}.

Our $H_{c2}(T)$ measurements in configuration with $H \parallel c$ for the most overdoped sample $x$=0.127 do not extend to low enough temperatures. However, bulk thermal conductivity measurements of Reid {\it et al.}, Ref.~\onlinecite{Reidcaxis}, made on the samples from the same batch, suggest $H_{c2,c}(0)$=10~T, as shown with star in the middle panel of Fig.~\ref{Hc2Comparison}. Compilation of the high temperature data from our study and of the low-temperature thermal conductivity data suggests that linear $H_{c2,c}(T)$ trend is indeed observed in $x$=0.127, the superconducting gap of which is characterized by the presence of nodes. For sample with $x$=0.108, we estimate $H_{c2,c}(0) \approx 30$~T.


\subsection{London penetration depth}


\begin{figure}[tb]
\begin{center}
\includegraphics[width=8cm]{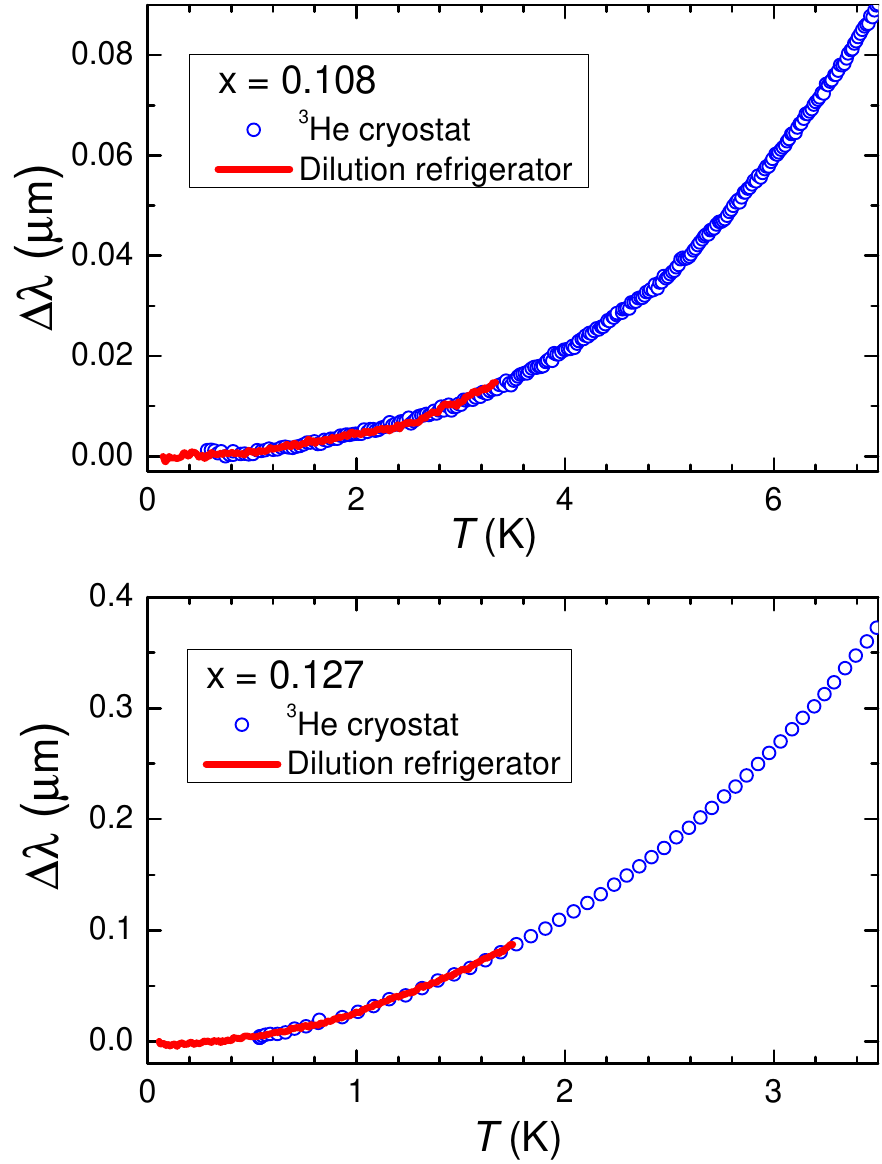}
\end{center}
\caption{(Color Online) Low temperature London penetration depth $\Delta$$\lambda$(T) for samples of Ba(Fe$_{1-x}$Co$_x$)$_2$As$_2$ with $x$=0.108 (top panel) and $x$=0.127 (bottom panel). Data were taken in both $^3$He-cryostat (down to $\sim$0.5~K, black curves) and a dilution refrigerator ($\sim$0.05~K$<T<$3~K, red curve), showing good matching between the data sets taken in two systems and the robustness of the power-law dependence.  }%
\label{pristinelinear}
\end{figure}


\begin{figure}[tb]
\begin{center}
\includegraphics[width=8cm]{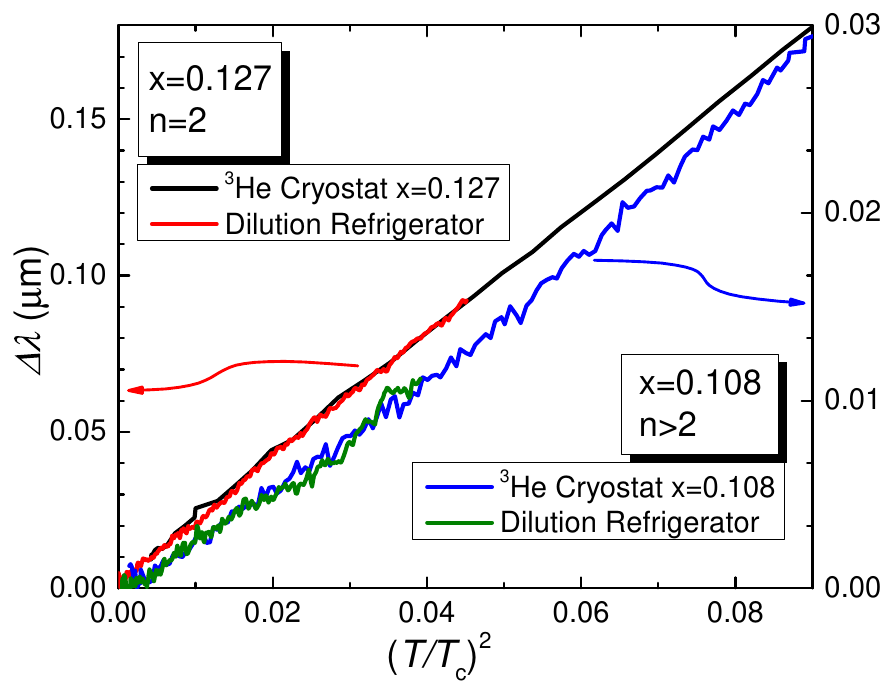}
\end{center}
\caption{(Color Online) Low temperature London penetration depth $\Delta$$\lambda$(T) measured in single crystals of Ba(Fe$_{1-x}$Co$_x$)$_2$As$_2$ with $x$=0.108 (green and blue curves) and $x$=0.127 (black and red curves) plotted vs. $(T/T_c)^2$ . Linear plot for $x$=0.127 shows that the dependence is very close to $T^2$, consistent with more detailed fitting analysis using floating fitting range, see Fig.~\ref{rangefitting} below. Clear deviations for sample $x$=0.108 suggest $n>2$.}%
\label{pristinesquare}
\end{figure}

Figure~\ref{pristinelinear} shows temperature-dependent variation of London penetration depth in pristine samples of BaCo122 with $x$=0.108 (top panel) and $x$=0.127 (bottom panel). Due to rather low $T_c\approx$8~K of the sample with $x$=0.127, measurements down to $\sim$0.5~K, the base temperature in our $^3$He setup, do not cover broad enough temperature range to give reliable power-law analysis. We extended the temperature range by taking the data in a dilution refrigerator down to $\sim$0.05~K, or $T_c/$160. The data sets taken in two systems perfectly match in the overlapping range 0.5 to 3.5~K providing support for the reliability of the measurements. It is clear from the inspection of the raw data, that the temperature variation of London penetration depth is much stronger than the exponential variation expected in full-gap superconductors. In fact the dependence is close to $T^2$, as shown in Fig.~\ref{pristinesquare}, in which the data for two compositions are plotted vs. $(T/T_c)^2$, which is similar to the earlier data by Gordon {\it et al.} \cite{Gordondoping}.
As can be seen from Fig.~\ref{pristinesquare}, the exponent $n$ is larger for the sample with closer to the optimal doping composition. Using a power-law fit over a temperature range up to $T_c/3$, we obtain $n$=2.5 for sample with $x$=0.108 and $n$=2.0 for $x$=0.127. These values and their change with doping follow general trend in iron-pnictides \cite{KyuilNaFeAs}. In BaCo122 this evolution is in line with the results of thermal conductivity \cite{TanatarPRL,Reidcaxis} and heat capacity \cite{BNC} studies.


\begin{figure}
\begin{center}
\includegraphics[width=8cm]{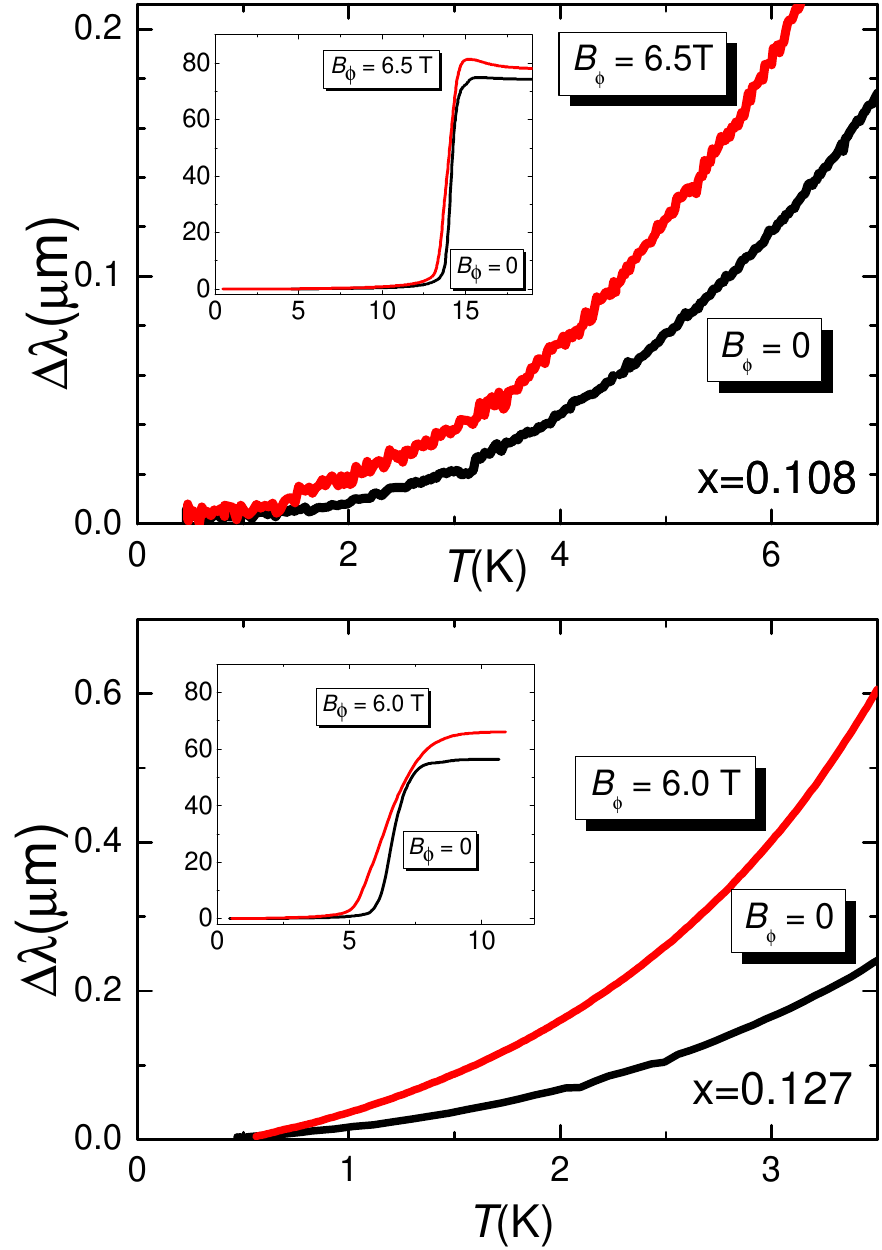}

\end{center}
\caption{(Color Online) Effect of heavy - ion irradiation on London penetration depth, $\Delta\lambda(T)$ in in samples with $x=0.108$ (top) and $x=0.127$ (bottom). Black curves show pristine samples, red - irradiated with matching fields of 6.5~T and 6~T respectively. Insets show variation of London penetration depth in the whole range up to $T_c$. }%
\label{irradiationlinear}
\end{figure}


\begin{figure}
\begin{center}
\includegraphics[width=8cm]{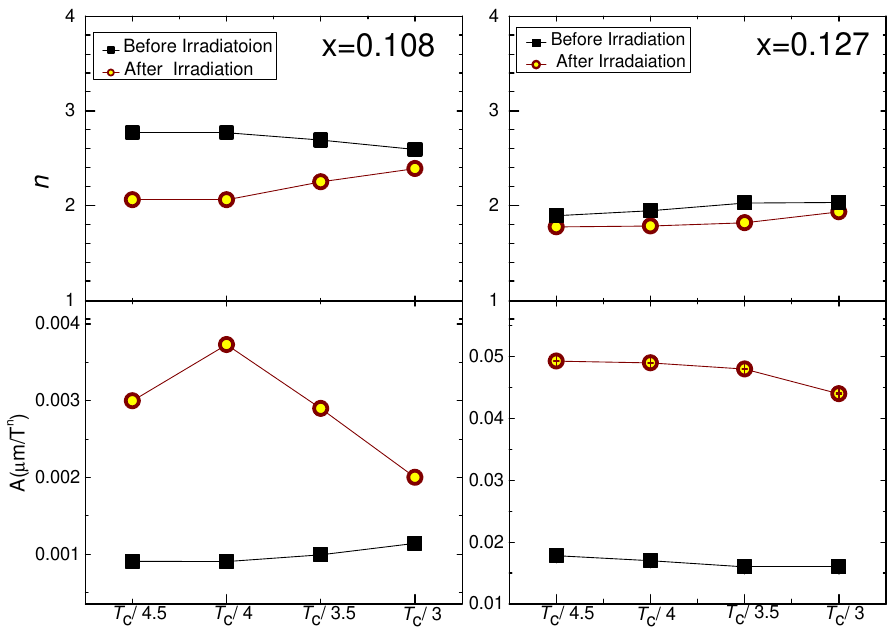}

\end{center}
\caption{(Color Online) Dependence of the fitting parameters, $n$ (top panels) and $A$ (bottom panels), of the power-law function, $\Delta \lambda =AT^n$, on the temperature of the high-temperature boundary of the fitting interval. Data are shown for pristine (black squares) and irradiated (yellow-brown circles) sample with $x$=0.108 (left) and $x$=0.127 (right).}%
\label{rangefitting}
\end{figure}

Figure~\ref{irradiationlinear} shows the London penetration depth from base temperature to $\sim T_c/3$ in the sample $x$=0.108 (top panel) before (black curve) and after 6.5T irradiation (red curve). Inset shows the data for the whole temperature range, revealing small but clear decrease of $T_c$. Irradiation significantly increases the total $\Delta \lambda (T)$ change from base temperature to $T_c/3$. The similar data for sample $x$=0.127 in pristine (black line) and 6~T irradiated (red line) states are shown in the bottom panel of Fig.~\ref{irradiationlinear}. The $T_c$ decrease in sample $x$=0.127 is larger than in sample $x$=0.108, and similarly, overall change in the penetration depth to $T_c/3$ is larger as well.

In a standard analysis of the penetration depth in single gap superconductors, the low-temperature asymptotic behavior is expected in the range from base temperature to roughly $T_c/3$, over the temperature range in which the superconducting gap itself can be considered as constant. This assumption is not valid for multi-band superconductors, in which case the high-temperature end of the fitting is determined by the smaller gap \cite{ProzorovROPP}. Since this ratio is $\textit{ a priori}$ unknown, we varied the high temperature range of the fit. We used a power-law function $\Delta \lambda (T)=AT^n$ and determined $n$ and $A$ as a function of the high-temperature end of the fitting range, always starting the fit at the base temperature. The results of this fitting analysis for pristine and irradiated samples are shown in Fig.~\ref{rangefitting}, for samples with $x$=0.108 (left column) and $x$=0.127 (right column). The top panels show the evolution of the exponent $n$ and the bottom panels show the evolution of the pre-factor $A$. The results of the fitting analysis, Fig.~\ref{rangefitting} indicate that for sample with $x$=0.108 the exponent $n$ weakly depends on the fitting range, changing from 2.7 to 2.6. In irradiated samples the exponent decreases to $n$=2.2 for $T_c/4.5$, and slightly increases to 2.3 for $T_c/3$. The decrease of the exponent with irradiation is not expected in either $s_{++}$ or $d-$wave states, but predicted for the $s_{\pm}$ pairing. The effect of irradiation is even more dramatic in a sample with $x$=0.127. Here the exponent in the pristine sample is $n$=2.0, a value possible to explain in both dirty $d-$wave and dirty $s_{\pm}$ scenarios \cite{HirschfeldGoldenfeld,GordonPRL}. In the former the exponent is expected to be insensitive to the increase of scattering, in the latter it is expected to decrease further down to about 1.6. As can be clearly seen, irradiation decreases $n$ to 1.8, suggesting an increase of anisotropy.
Simultaneously, the prefactor in these samples also increases after irradiation, clearly showing the appearance of excess quasi-particles due to additional in-gap density of states induced by pair-breaking scattering.


\section{Conclusions}

In conclusion, we find that the temperature-dependent London penetration depth in overdoped samples of BaCo122, is best fit with the power-law with the exponent $n$ decreasing with $x$ towards the overdoped edge of the superconducting dome, confirming increasing gap anisotropy. The exponent $n$ also decreases after heavy-ion irradiation introducing additional scattering. This observation is in line with the expectations for $s_{\pm}$ pairing state with accidental nodes but contradicts those for $s_{++}$ state. It suggests that $s_{\pm}$ pairing state is universal over the whole doping range in electron-doped BaCo122.

Considering our resistively measured $H_{c2,c}(T)$ together with the results of the previous thermal conductivity studies, we find that the $H_{c2,c}(T)$ dependence in sample $x$=0.127, at the very dome edge with nodes in the superconducting gap, is very close to $T-$linear. This observation is in line with our finding of the link between the superconducting gap anisotropy and the anomalous $T-$linear dependence of $H_{c2,c}$. It is suggestive that the feature may be universal in iron-arsenide superconductors.

\section{Acknowledgments}
Work at Ames Laboratory was supported by the U.S. Department of Energy, Office of Basic Energy Sciences, Division of Materials Sciences and Engineering under contract No. DE-AC02-07CH11358.
Work at Argonne was supported by the Center for Emergent Superconductivity, an Energy
Frontier Research Center funded by the US Department of
Energy, Office of Science, Office of Basic Energy Sciences
under Award No. DE-AC0298CH1088.
Work at the National High
Magnetic Field Laboratory is supported by the NSF Cooperative
Agreement No. DMR0654118 and by the state of Florida.


\end{document}